# External Control of Magnetism in Semiconductors at High Temperature (~100K)

**One-sentence summary**: We report the electrical and optical control of ferromagnetism in semiconductor heterostructures at high temperatures of 100-117 K, by manipulating the electric-field-induced hole concentration and the photo-generated hole-spins, respectively.

Ahsan M. Nazmul,[1,2] S. Kobayashi,[1] S. Sugahara,[1,2] and M. Tanaka[1,2*]

1. Department of Electronic Engineering, The University of Tokyo, 7-3-1 Hongo, Bunkyo-ku, Tokyo 113-8656, Japan.

2. PRESTO, Japan Science & Technology Corporation, 4-1-8 Honcho, Kawaguchi, Saitama 332-0012, Japan.

*To whom correspondence should be addressed. masaaki@ee.t.u-tokyo.ac.jp

**Abstract:** We demonstrate the electrical and optical control of ferromagnetism in semiconductor heterostructures at high temperatures of 100–117 K. The heterostructures consist of Mn delta ($\delta$)-doped GaAs and *p*-type AlGaAs, where the overlap of the hole wavefunction with the Mn $\delta$-doping profile leads to high ferromagnetic transition temperature $T_C$ of over 100 K (max. 172 K). We are able to isothermally change the paramagnetic state to the ferromagnetic state and *vice versa*, by applying a gate electric-field or by light irradiation. The large modulation of $T_C$ ($\Delta T_C$ ~15 K) at high temperatures (> ~100 K) demonstrated here may pave the way to functional device applications compatible with the present semiconductor technology.

While the bandgap engineering and wavefunction engineering are so far limited to *nonmagnetic* semiconductor heterostructures, broadening their frontiers to *magnetic* heterostructures is expected to afford a new spin degree of freedom in designing the functions of semiconductors. Utilizing the cooperative phenomena of magnetic spins and carriers in *magnetic* semiconductor heterostructures opens the way to control the ferromagnetism by changing the carrier characteristics such as the concentration, spin, and wavefunction. The external control of ferromagnetism by changing the carrier concentration using a gate electric field has recently been shown in a few magnetic semiconductor materials of the group-II-VI-based CdMnTe (*1*), -III-V-based InMnAs (*2*), and -IV-based MnGe (*3*), at the operation temperature of 1.3 K, 22 K, and 50 K, respectively. On the other hand, the manipulation of Mn and carrier spins by using circularly polarized light has been reported since the 1980's in II-VI-based magnetic semiconductor materials at 4.2 K (*4*) – 5.8 K (*5*), and very recently in III-V based materials at 4.2 K (*6*). While these experimental results provide fertile ground for fundamental studies, the device application is yet to see any success because of the low ferromagnetic transition temperature (or Curie Temperature $T_C$) far below room temperature.

There grew more attention to III-V-based magnetic alloy semiconductor materials such as (InMn)As and (GaMn)As (*7-9*) since the 1990's, because their semiconductor hosts, especially GaAs, are already used for both electronic and optical devices, such as high-speed transistors and semiconductor lasers. Despite higher $T_C$ (110-160 K) in GaMnAs (*10-13*) than most of the other group-II-VI, -III-V, and -IV based magnetic semiconductors, the external control of the ferromagnetism in the GaAs-based materials using electric field or light remains to be realized, probably due to the underlying difficulty in the control of a large number of holes of the order of $10^{19}$-$10^{20}$ cm$^{-3}$ required to align the magnetic Mn spins in the material.

In this article, we take a different approach to control the ferromagnetism using both electric field and light at high temperature over 100 K, in which the semiconductor bandgap-engineering is utilized for the external control of ferromagnetism. We have grown Mn delta($\delta$)-doped GaAs / Be-doped $Al_{0.3}Ga_{0.7}As$ heterostructures by molecular-beam epitaxy (MBE), as shown in Fig. 1A, which is a *p*-type selectively doped heterostructure (*p*-SDHS) with a Mn $\delta$-doped layer in the hole-conduction channel. $\delta$-doping of magnetic atoms (Mn) is used to provide a high concentration of local magnetic moments in a $\delta$-function-like Mn profile along the growth direction. In the heterostructures of Fig. 1A, holes are provided from the Be-doped *p*-type AlGaAs layer to the Mn $\delta$-doped GaAs layer, and the effective overlap of the local Mn spins and the 2-dimensional hole gas (2DHG) wavefunction leads to the ferromagnetic order with $T_C$ up to 172 K (*14*), the highest among the $T_C$ values ever reported in III-V (InAs, GaAs) based magnetic semiconductors. A suitable *p*-SDHS with Mn $\delta$-doping is expected to allow external control of ferromagnetism at high temperature, through electrostatically induced holes by gate electric field or photo-induced hole-spins by circularly polarized light. As predicted by the Zener model mean field calculation (*15*), the high local-density of Mn spins and sufficient hole concentration will give rise to ferromagnetic order with high $T_C$ through the *p-d* exchange interaction between the Mn spins and the holes.

The sample structure examined here is shown in Fig. 1A and 1B. The Mn concentration (or "coverage") $\theta_{Mn}$ in the $\delta$-doped layer was 0.25 monolayer (ML), where $\theta_{Mn}$ = 1 ML corresponds to a sheet concentration of $6.3 \times 10^{14}$ $cm^{-2}$ (*16-17*). The MBE growth sequence from the bottom is; a 200 nm-thick undoped GaAs buffer, a 300 nm-thick undoped $Al_{0.3}Ga_{0.7}As$ layer, and a 30 nm-thick Be-doped $Al_{0.3}Ga_{0.7}As$ layer (Be concentration = $1.8 \times 10^{18}$ $cm^{-3}$) were grown at a substrate temperature $T_s$ = 600°C on a semi-insulating (SI) (001) GaAs substrate. Then followed a 0.25 ML Mn $\delta$-doped layer and a 20 nm-thick undoped GaAs cap layer grown at $T_s$ = 300°C. The relatively thick (300 nm) undoped $Al_{0.3}Ga_{0.7}As$ layer was inserted between the *p*-AlGaAs layer and the undoped GaAs buffer layer to avoid the accumulation of holes at the AlGaAs/GaAs interface on the substrate side. After the MBE growth, a 195 nm-thick $SiO_2$ gate insulation layer and a 100 nm-thick Al electrode were deposited on the sample surface to fabricate a field effect transistor (FET) structure. The FET structure with a channel width of 50 μm and a length of 200 μm, schematically shown in Fig. 1B, was defined in a Hall bar configuration using photolithography and chemical etching.

We measured the anomalous Hall effect (AHE) (*18*) to study the magnetic properties. This is a very sensitive method to characterize the magnetization of ultra-thin films and heterostructures of quasi-two dimensional carrier systems, for which the bulk magnetization measurement is difficult. By taking into account the AHE contribution to the Hall resistance of a quasi-two dimensional magnetic material, the sheet Hall resistance $R_H$ is expressed as (*14*)

$$R_H = R_O B + R_S M \qquad (1)$$

Here, $R_O$ is the ordinary sheet Hall coefficient, $B$ is the applied magnetic field, $R_S$ is the anomalous sheet Hall coefficient and $M$ is the perpendicular component of magnetization (or "perpendicular magnetization") of the sample. In the ferromagnetic phase at low temperature, the Hall effect is dominated by the second term, that is AHE, in Eq. 1. Hence, $R_H$ can be approximately expressed as

$$R_H \approx R_S M \qquad (2)$$

In other words, the Hall resistance $R_H$ is proportional to the magnetization $M$. Therefore, the Hall measurement will give a good account of the magnetization of our ultra-thin magnetic heterostructures.

The change of the magnetic properties in the *p*-SDHS of Fig. 1A under zero and finite gate biases is depicted in Fig. 2A-2B by using the energy band diagrams. In the *p*-SDHS, two-dimensional hole gas (2DHG) is formed at the GaAs/*p*-AlGaAs heterointerface, where there exists a locally high concentration δ-doped Mn spins. The *p*-SDHS in a FET structure allows the external control of the 2DHG concentration $p_{sheet}$, thus its magnetic state, by using a gate bias $V_G$. As shown in Fig. 2A, $p_{sheet}$ is low at $V_G = 0$, which is not sufficient to realize the ferromagnetic order among the Mn spins, thus the heterostructure is paramagnetic. Under a negative gate bias applied to the Al gate ($V_G < 0$), holes are electrostatically induced in the Mn δ-doped GaAs layer, as shown in Fig. 2B. This increase of $p_{sheet}$ leads to isothermal phase transition to the ferromagnetic state owing to the enhanced hole-Mn interaction through the 2DHG wavefunction overlapping.

Figure 2C and 2D show the Hall resistance ($R_H$) loops of the heterostructure of Fig. 1A as a function of magnetic field (*B*) under different gate biases ($V_G = 0$ V, -15 V, and -18 V) measured at 115 K and 117 K, respectively. In Fig. 2C and 2D, the Hall characteristics changed from a linear trace to a clear ferromagnetic hysteresis loop with the application of negative gate biases, indicating the phase transition from the paramagnetic state to the ferromagnetic state. This indicates that at negative gate biases the increased accumulation of holes in the Mn δ-doped channel induced long-range ferromagnetic ordering among the Mn spins. In Fig. 2C and 2D, one can see ferromagnetic hysteresis plus a positive linear component even at $V_G = -15$ V and -18 V, which is probably caused by the parasitic parallel conduction of holes in the *p*-AlGaAs layer or in the GaAs buffer layer or the cap layer (*19*).

We measured the temperature dependence of the spontaneous Hall resistance (corresponding to the spontaneous magnetization), which was derived by subtracting the linear component from the $R_H - B$ characteristics at different gate biases, and thereby, $T_C$ at $V_G$ = 0 V, -15 V, and -18 V was estimated to be 105 K, 115 K, and ~120K, respectively. Here, the field-effect enhanced change in the transition temperature $\Delta T_C$ was as large as 10 K and 15 K for $V_G$ = -15 V and -18 V, respectively. The modulated temperature range of $T_C$ (105 K ~ 120 K) in our Mn δ-doped GaAs heterostructure is much higher and wider compared with that ($T_C$ = 26 K ~ 27 K, $\Delta T_C$ = 1 K) of another III-V magnetic semiconductor of InMnAs, and the required gate voltage (-15 V ~ -17 V) was substantially smaller than that (-125 V) in InMnAs (*2*). The present highly-modulated $T_C$ by the gate electric-field is attributed both to the locally high concentration of Mn spins in the δ-doping profile and to the easy modulation of hole concentration in our heterostructure.

In the following, we discuss more quantitatively the change of $T_C$ and sheet hole concentration $p_{sheet}$ induced by the gate voltage. First, we estimated the $p_{sheet}$ to be $6.3 \times 10^{11}$ cm$^{-2}$ at $V_G$ = 0, by ruling out the AHE contribution and using the Curie Weiss fitting. Details of the expressions and fitting procedures are described elsewhere (*14*). From the capacitance-voltage (C-V) measurement of the gate-insulator of the FET structure at 115 K, it is estimated that the negative gate biases of $V_G$ = -15 V and -18 V increase the $p_{sheet}$ by $1.5 \times 10^{12}$ cm$^{-2}$ and $2.1 \times 10^{12}$ cm$^{-2}$, respectively. Hence, the $p_{sheet}$ at −15 V and −18 V becomes $2.1 \times 10^{12}$ cm$^{-2}$ and $2.7 \times 10^{12}$ cm$^{-2}$, respectively, assuming that the mobility is constant. Using the mean-field theory of the Zener model for magnetic alloy semiconductors (*15*), the calculated change of $T_C$ ($\Delta T_C$) for the increased $p_{sheet}$ at $V_G$ = -15 and -18 V were 11 K – 14 K, respectively. In this calculation, we set the Mn content $x$ = 0.125 for $\theta_{Mn}$ = 0.25 ML (since we defined the local Mn content as $\theta_{Mn}$ /2 ML, where 2 ML is the width of the Mn distribution along the growth direction estimated by transmission electron microscopy) (*17*). The *p-d* exchange energy $N_0\beta$ used in this calculation was −1.2 eV, which is the value of our MBE-grown (Ga,Mn)As alloy sample estimated by photoemission studies (*20*). This theoretically estimated $\Delta T_C$ of 11 K - 14 K is in good agreement with the experimentally observed $\Delta T_C$ = 10 K - 15 K for $V_G$ = -15 V ~ -18 V.

The above heterostructure also gives another opportunity to externally control the ferromagnetism using light with photon energy greater than the semiconductor band-gap ($E_g$). It is expected that photo-generated holes, produced by the irradiation of light on the heterostructure of Fig. 1A, are preferentially populated in the triangular potential well at the GaAs/(Al,Ga)As interface whereas photo-generated electrons are driven to the surface (see the schematic band diagram in Fig. 3A), leading to the enhancement of the ferromagnetic order among the Mn spins in the Mn δ-doped GaAs layer. A significant advantage of light irradiation compared with the gate-electric field is that the spin-polarity of the photo-generated holes can be controlled by the circular polarization of the incident light. This may afford a method for the direct control of the magnetization in the heterostructures by changing the polarity of hole-spins. Hence, we measured AHE, thus magnetization, both in the dark and under the light of right ($\sigma^+$), left ($\sigma^-$) circular polarization, and linear polarization (LP) using a He-Ne laser whose photon energy is 1.96 eV and light power intensity is 8 mW/cm$^2$. The $\sigma^+$ (or $\sigma^-$) light generates spin-polarized electrons and holes with the polarization $P(=\frac{n_\uparrow - n_\downarrow}{n_\uparrow + n_\downarrow})$ of 50%, where $n_\uparrow$ and $n_\downarrow$ are the number of carriers with up spin and down spin, respectively. The growth and structural parameters of the sample were the same as the structure of Fig. 1A, except that the Mn concentration $\theta_{Mn}$ was 0.30 ML. The magnetotransport properties were measured in a Hall bar configuration, whose geometry is the same as Fig. 1B, except that no insulation layer and no metal gate were deposited on the surface.

Figure 3B shows the Hall resistance ($R_H$) loops of the heterostructure measured at 80 K under different polarizations of light, and also in the dark. While the $R_H$-$B$ characteristics in the dark and under LP-light remained unchanged, $R_H$ changed in the opposite manner for the $\sigma^+$ and $\sigma^-$ light. The magnetization $M$, which is proportional to $R_H$ (see Eq. 2), increased under $\sigma^+$ light, whereas it decreased under $\sigma^-$ light. These results indicate that the photo-induced magnetization is caused by the hole-spins generated by the circularly polarized light. These observations are consistent with those in GaMnAs alloy thin films (*6*), suggesting that a modulated *p-d* exchange mechanism between the Mn-spins and hole-spins may be responsible for the change in magnetization.

The Hall loops of the heterostructure at 95 K and 100 K under the magnetic field of –1 T ~ +1 T are shown in Fig. 4 measured in the dark and under $\sigma^+$ light. The remnant magnetization (the remanence in the Hall loop) measured at different temperatures led to $T_C$ = 105 K under $\sigma^+$ light, whereas $T_C$ of the same sample in the dark condition was 95 K. Compared with the results recently reported in GaMnAs alloy thin films (6), where the change of the magnetization was observed at very low temperature of 4.2 K using a much stronger light power of 600 mW/cm$^2$, the magnetization and $T_C$ in our Mn δ-doped GaAs heterostructure were significantly changed by $\sigma^+$-light-induced hole-spins at higher temperatures in a wide range ($T_C$ was increase from 95 K to 105 K) with a much lower light power (8 mW/cm$^2$). By ruling out the AHE contribution using the Curie Weiss fitting (*14*), we estimated the $p_{sheet}$ in the dark to be $1.7 \times 10^{12}$ cm$^{-2}$. Assuming a constant mobility, the $p_{sheet}$ under $\sigma^+$ light at 95 K increased by $\Delta p_{sheet} = 1.3 \times 10^{12}$ cm$^{-2}$ up to $3.0 \times 10^{12}$ cm$^{-2}$. Under $\sigma^+$ light at 95 K, $R_H$ (thus the magnetization) increased by a factor of 0.45 compared with that in the dark when the applied magnetic field was 1 T (see Fig. 4A). This means that the spins of nearly 65 Mn atoms were rotated toward the perpendicular direction, when one hole-spin is generated by the $\sigma^+$ light in the heterostructure.

The large modulation of $T_C$ ($\Delta T_C$ = 10 K ~ 15 K) using both gate electric field and light irradiation at high operation temperatures (~ 100 K, which is well over the liquid nitrogen temperature of 77 K) presented here in the III-V-based magnetic heterostructures can lead to functional device applications combining "spintronics" and "optoelectronics" compatible with the present semiconductor technology. Although the modulated $T_C$ range is still below room temperature, more suitably fabricated magnetic semiconductor heterostructures with proper preparation conditions could further increase $T_C$ (*14*) in line with the prediction of the Zener-model theory (*15*).

**Figure Captions**

**Fig. 1. (A)** Illustration of a 0.25 ML Mn δ-doped GaAs / Be-doped *p*-AlGaAs heterostructure grown by molecular beam epitaxy on a semi-insulating (SI) (001) GaAs substrate. The structure is a *p*-type selectively doped heterostructure (*p*-SDHS) with Mn δ-doping in the 2-dimensional hole gas (2DHG) channel. After the MBE growth, a 195 nm-thick $SiO_2$ insulation layer and a 100 nm-thick Al gate electrode were deposited on the surface. **(B)** Illustration of a Hall bar with a gate electrode defined by photolithography and chemical etching. The channel width and length are 50 μm and 200 μm, respectively.

**Fig. 2.** Valence band diagrams of the heterostructure of Fig. 1A at zero gate bias **(A)**, and at a finite negative gate bias **(B)**, respectively, with hole energy along the vertical axis. $E_V$ and $E_F$ are the valence band edge and the Fermi energy, respectively. $z$ is the growth direction. The GaAs separation layer thickness $d_s$, a measure to control the ferromagnetic ordering with the variable overlap of the 2DHG wavefunction and the Mn δ-doped profile (*14*), was set to 0 nm for the samples examined here. Hall resistance ($R_H$) loops of the sample measured at 115 K **(C)**, and at 117 K **(D)**, respectively, are shown as a function of magnetic field (*B*) for different gate biases ($V_G = 0$, -15 and -18 V).

**Fig. 3. (A)** Band profile of the Mn δ-doped GaAs / *p*-type AlGaAs heterostructure (not drawn to the scale), while the sample is irradiated with light. Photo-generated holes preferentially flow into the triangular potential well at the Mn δ-doped GaAs/AlGaAs interface contributing to the enhancement of the ferromagnetic order, whereas photo-generated electrons are driven to the surface. **(B)** Hall resistance ($R_H \propto M$) characteristics of the heterostructure measured at 80 K under light with linear polarization (LP), right ($\sigma^+$) and left ($\sigma^-$) circular polarizations, and also in the dark.

**Fig. 4**. Hall resistance ($R_H \propto M$) loops of the heterostructure sample in the dark and under light measured at 95 K **(A)**, and at 100 K **(B)**. The enlarged figures of the Hall loops of (A) and (B) at low magnetic fields (-0.2 T ~ 0.2 T) are shown in **(C)** and **(D)**.

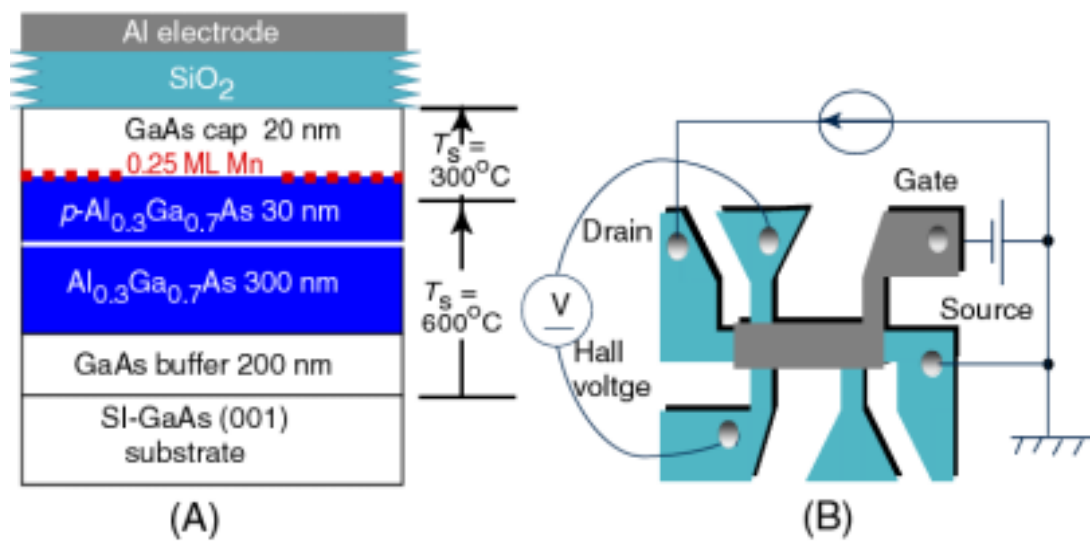

Fig. 1     Nazmul *et al*.

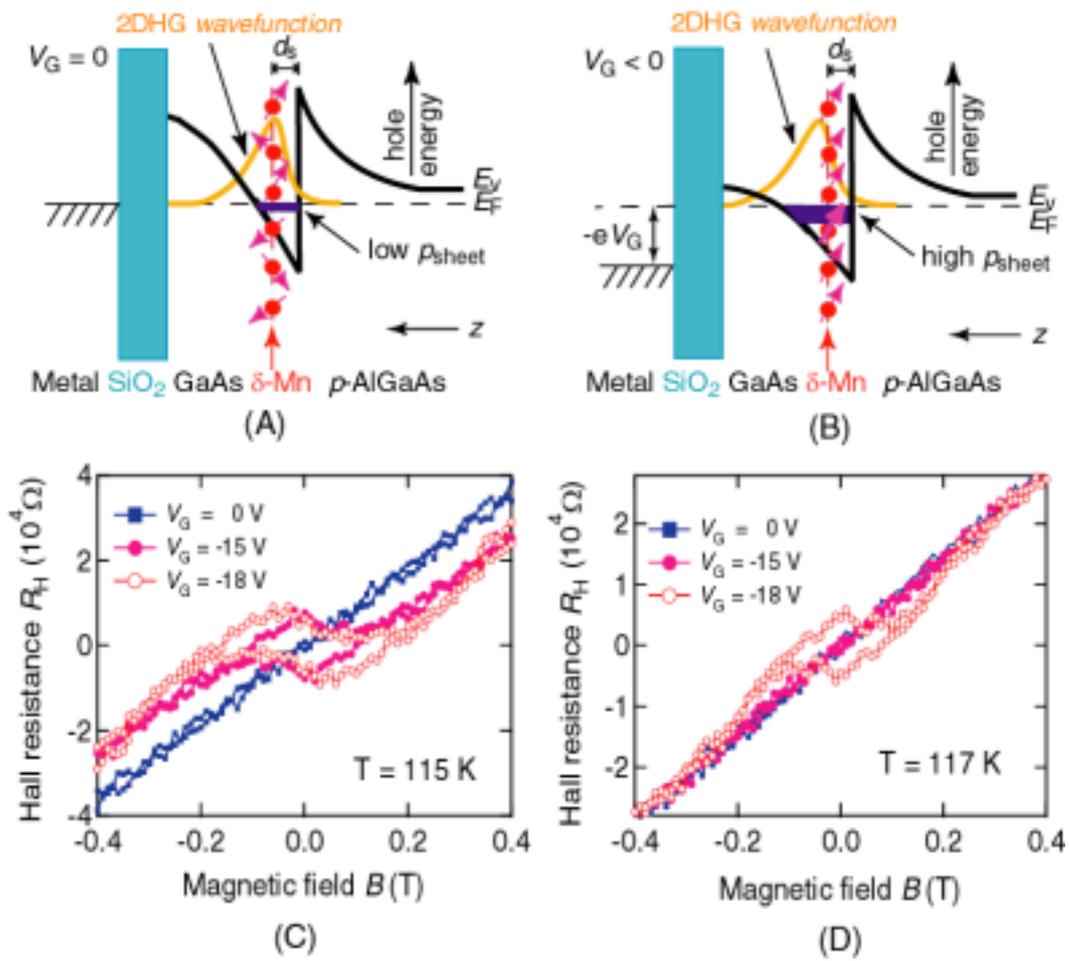

Fig. 2 Nazmul *et al*.

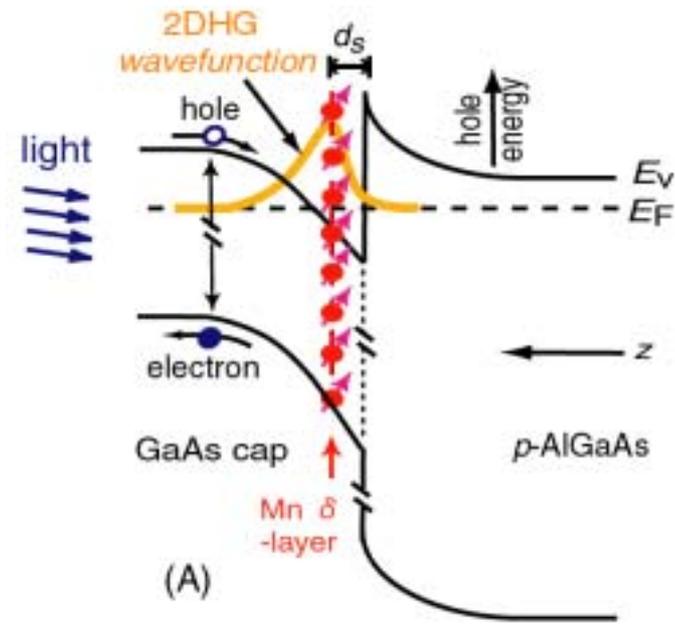

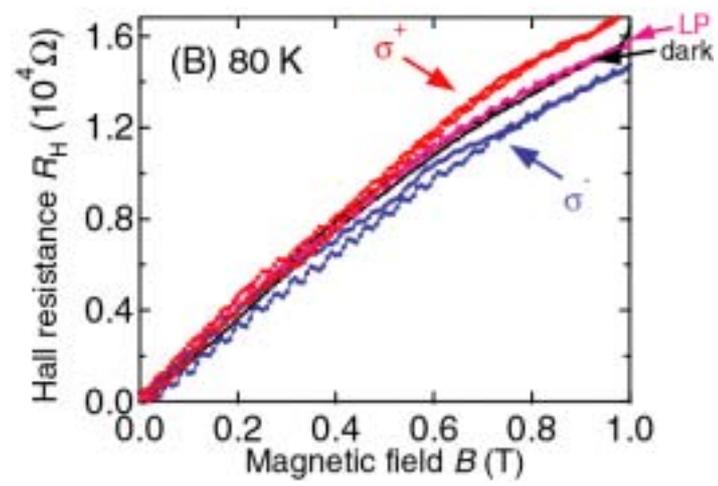

Fig. 3　　　Nazmul *et al*.

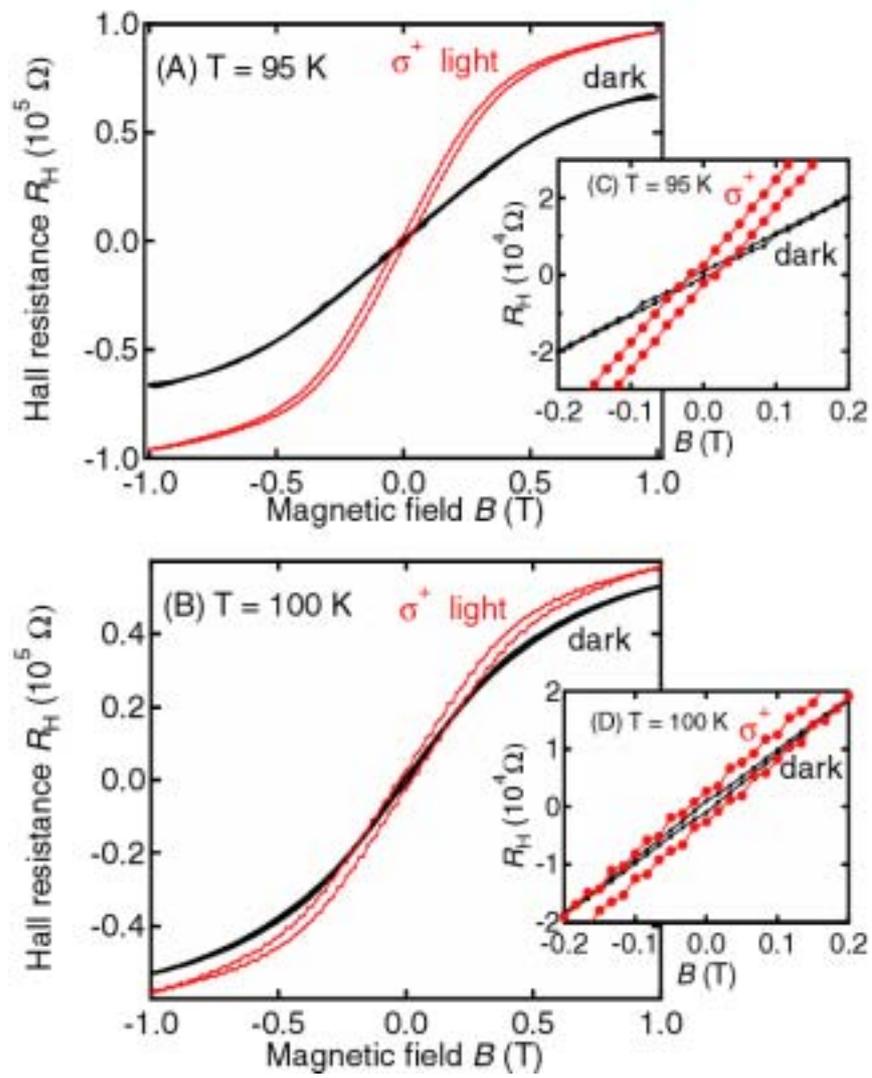

Fig. 4    Nazmul *et al*.